\documentclass{article}

\usepackage[preprint]{neurips_2024}

\usepackage[utf8]{inputenc} % allow utf-8 input
\usepackage[T1]{fontenc}    % use 8-bit T1 fonts
\usepackage{hyperref}       % hyperlinks
\usepackage{url}            % simple URL typesetting
\usepackage{booktabs}       % professional-quality tables
\usepackage{amsfonts}       % blackboard math symbols
\usepackage{nicefrac}       % compact symbols for 1/2, etc.
\usepackage{microtype}      % microtypography
\usepackage{xcolor}         % colors

\usepackage{algorithm}
\usepackage{algorithmic}
\usepackage{amsmath}
\usepackage{amssymb}
\usepackage{dsfont}
\usepackage{graphicx}
\usepackage{kotex}

\title{Hypernetwork-Based Approach for Optimal Composition Design in Partially Controlled Multi-Agent Systems}

\author{%
    Kyeonghyeon Park\\
    Korea Advanced Institute of Science and Technology\\
    \texttt{kyeonghyeon.park@kaist.ac.kr} \\
    \And
    David Molina Concha\\
    University of Toronto\\
    \texttt{damolina@mie.utoronto.ca}\\
    \And
    Hyun-Rok Lee\\
    Inha University\\
    \texttt{hyunrok.lee@inha.ac.kr}\\
    \And
    Chi-Guhn Lee\\
    University of Toronto\\
    \texttt{cglee@mie.utoronto.ca}\\
    \And
    Taesik Lee\\
    Korea Advanced Institute of Science and Technology\\
    \texttt{taesik.lee@kaist.edu}\\
}

\begin{document}

\maketitle

\begin{abstract}

Partially Controlled Multi-Agent Systems (PCMAS) are comprised of controllable agents, managed by a system designer, and uncontrollable agents, operating autonomously. This study addresses an optimal composition design problem in PCMAS, which involves the system designer's problem, determining the optimal number and policies of controllable agents, and the uncontrollable agents' problem, identifying their best-response policies. Solving this bi-level optimization problem is computationally intensive, as it requires repeatedly solving multi-agent reinforcement learning problems under various compositions for both types of agents. To address these challenges, we propose a novel hypernetwork-based framework that jointly optimizes the system's composition and agent policies. Unlike traditional methods that train separate policy networks for each composition, the proposed framework generates policies for both controllable and uncontrollable agents through a unified hypernetwork. This approach enables efficient information sharing across similar configurations, thereby reducing computational overhead. Additional improvements are achieved by incorporating reward parameter optimization and mean action networks. Using real-world New York City taxi data, we demonstrate that our framework outperforms existing methods in approximating equilibrium policies. Our experimental results show significant improvements in key performance metrics, such as order response rate and served demand, highlighting the practical utility of controlling agents and their potential to enhance decision-making in PCMAS.

\end{abstract}

\section{Introduction}

Complex systems composed of autonomous agents, each aiming to maximize individual objectives, often result in suboptimal outcomes for the overall system due to misalignment between individual agent goals and system objectives \citep{dubey1986inefficiency}.

The concept of Partially Controlled Multi-Agent Systems (PCMAS) has been introduced \citep{brafman1996partially} to address this problem. PCMAS are systems in which only a subset of agents is directly controlled by a system designer. In such systems, the designer strategically selects and manages these controllable agents to improve overall system performance, considering practical constraints such as agents' autonomy, cost, and scalability. This paradigm has become increasingly relevant in diverse domains, including smart grids, healthcare, and traffic management. For example, in smart grids, contracts balance energy production and consumption across stakeholders \citep{le2020ethical}; in healthcare, public hospitals complement private ones to optimize social welfare \citep{xue2023relationship}; and in traffic management, autonomous vehicles improve traffic flow or mitigate demand loss caused by human-driven vehicles \citep{xie2023two,di2021survey,lazar2021learning}.

Within the PCMAS framework, a central problem is to determine the number (or fraction) of controllable agents and their policies in the target multi-agent system, which we refer to as a PCMAS composition design problem. Conceptually, it can be formulated as a bi-level optimization problem. At the upper level, the system designer's decision is the number of controllable agents to maximize system performance. At the lower level, given an agent composition determined by the upper-level decision, optimal cooperative policies for the controllable agents are derived. In doing so, the best-response policies for uncontrollable agents must also be computed concurrently. By solving the lower-level problem, the system outcome is estimated for a given composition and subsequently fed back to the upper-level problem. This problem is computationally intensive because it requires repeatedly solving the lower-level multi-agent reinforcement learning (MARL) tasks across various compositions.

Although previous studies have shown some progress, they leave large room for improvement. Approaches based on Bayesian optimization (BO) have shown promise in reducing the computational burden at the upper level \citep{molina2024bayesian,molina2024algorithmic,shou2020reward}, but are subject to an inherent difficulty as they require re-learning equilibrium policies in MARL for every new composition. Adaptive incentive design (AID) has attempted to reduce computation time by combining designer and agent decision-making into a single loop; however, it often faces instability issues \citep{yang2022adaptive}. Most notably, while prior research on mixed autonomy has explored policy identification for different types of agents, it treats agent composition as a given parameter, rather than addressing it as a system-level optimization problem \citep{xie2023two}.

To address these challenges, we propose a novel composition design framework based on hypernetworks. Unlike traditional approaches that train separate networks for each system composition, our framework employs unified training across all compositions, thereby enabling efficient policy generation for both controllable and uncontrollable agents. This approach reduces computational burden and enhances information sharing across similar configurations. By incorporating the reward parameters of controllable agents, our framework effectively optimizes their policies. Furthermore, we integrate a network that predicts mean actions to improve scalability in large-scale systems. To evaluate the performance of our framework, we constructed an environment using real-world New York City (NYC) taxi data. Numerical results demonstrate that our framework approximates equilibrium solutions and outperforms existing methods. They also reveal that utilizing controllable agents is beneficial only within certain ranges of system size (i.e., the total number of agents), offering valuable insights for system design. 

Our framework is the first to employ  hypernetworks for generating multi-agent policies in system design, marking a significant advancement in this field.  Overall, the main contributions of this work are fourfold: (1) We propose a novel framework for PCMAS composition design based on hypernetworks, which enables unified training across all system compositions, thereby improving computational efficiency and information sharing. (2) The framework integrates controllable agents’ reward parameters to enhance policy optimization and incorporates a mean action prediction network to improve scalability for applications in large-scale systems. (3) Using real-world NYC taxi data, we demonstrate that our framework effectively approximates equilibrium solutions and outperforms existing methods. (4) Our study offers insights into effective utilization of controllable agents under varying system conditions and highlights the framework's flexibility by optimizing diverse objective functions.

\section{Related work}

\paragraph{System design} 

The composition design problem in PCMAS aligns with contract design \citep{dutting2023multi} and algorithmic mechanism design \citep{nisan2001algorithmic}, focusing on optimizing controllable agents' composition and operation to enhance system performance. Recent works on fleet design highlight fleet size's impact on system efficiency \citep{molina2024bayesian,barrios2014fleet,cabrera2014fleet}. While these works introduced a BO framework for robot fleet design, they often face substantial computational challenges. Similarly, research in PCMAS's mixed-autonomy domain has proposed frameworks for learning policies applicable to both human-driven and autonomous vehicles \citep{xie2023two}. However, these studies overlook system-level optimization from a compositional perspective, equilibrium considerations, and convergence stability concerns. Concurrently, advancements in reward design have explored single-loop approaches integrating reward design with agent learning \citep{yang2022adaptive,li2020end,guresti2023iq}. Despite progress, these methods face limitations such as restrictive theoretical assumptions, stability issues, and the lack of equilibrium-based solutions.

\paragraph{Computational efficiency}

On the other hand, research has increasingly focused on addressing the computational efficiency of MARL. Mean-field reinforcement learning (MFRL) simplifies interactions by approximating other agents' actions as a mean action \citep{yang2018mean}. However, MFRL often relies on historical mean actions to compute current actions, potentially causing delays in agent interactions. To overcome this, recent studies have developed methods to predict mean actions more effectively, which have been incorporated into our work \citep{zhou2020multi,li2024beyond}. Meanwhile, hypernetworks, which are specialized neural networks designed to generate weights for other networks, enable dynamic adaptation, improved generalization, and reduced trainable parameters by leveraging shared structures across tasks \citep{chauhan2023brief}. They have been applied in domains such as AutoML, zero-shot learning, multitasking, and RL. For instance, in computer vision, hypernetworks dynamically adjust feature extraction layers. In RL, they have been used in QMIX \citep{rashid2020monotonic} to mix individual Q-values and explored as alternatives to standard Q-function architectures \citep{sarafian2021recomposing}. Recently, population-size-aware policy optimization (PAPO) was introduced to bridge finite-agent and infinite-agent game theory in mean-field games \citep{li2023population}. However, its reliance on single-agent RL limited equilibrium guarantees in multi-agent systems. While hypernetworks offer strong generalization capabilities, challenges like potential accuracy losses remain only partially addressed. Despite their widespread use across fields, applying hypernetworks to mechanism design in systems with non-cooperative agents remains an open research area.

\section{Preliminaries}

\subsection{The composition design problem}

The composition design problem in PCMAS involves determining the optimal number of controllable agents ($c$-agents) and their corresponding cooperative policies, while considering the autonomy of uncontrollable agents ($u$-agents) and adhering to constraints such as budget or system efficiency. Formally, the set of possible PCMAS, denoted as $\mathcal{M}^\mathbf{w}$, is defined as the collection of Markov games (MGs) induced by all possible system compositions $\mathbf{w}$:
\[
\mathcal{M}^\mathbf{w}\equiv\{<\mathcal{N}, \mathcal{S}, \mathcal{A}, \mathcal{P}, \mathcal{R}, \gamma>|\mathcal{N}=\mathcal{N}_u \cup \mathcal{N}_c\},
\]
where $\mathbf{w}=\{N_u, N_c\}$, with $N_k$ representing the number of $k$-agents for $k \in \{u, c\}$. Here, $\mathcal{N}_k=\{1,\cdots,N_k\}$ denotes the set of $k$-agents and $\mathcal{N}$ represents the complete set of agents. The set of states is denoted by $\mathcal{S}$, and the (joint) action space is defined as $\mathcal{A}=\mathcal{A}_u\times\mathcal{A}_c=\Pi_{i_u\in\mathcal{N}_u}\mathcal{A}^{i_u}\times\Pi_{i_c\in\mathcal{N}_c}\mathcal{A}^{i_c}$, where each agent $i_k$ selects actions from their respective action spaces. Lastly, the state transition function is denoted by $\mathcal{P}: \mathcal{S} \times \mathcal{A} \times \mathcal{S} \mapsto [0, 1]$, the reward function by $\mathcal{R}: \mathcal{S} \times \mathcal{A} \times \mathcal{S} \mapsto \mathbb{R}$, and the discount factor by $\gamma$.

Each $u$-agent $i_u \in \mathcal{N}_u$ aims to find a policy $\pi_u^{i_u}$ that maximizes its expected return. The system designer, on the other hand, seeks to contract certain agents at a cost to control them toward achieving system-level goals and improving efficiency. The key challenge is determining the optimal number of $c$-agents $N_c^*$ and their cooperative policies $\pi_c$ to meet the system's objectives. While adding more $c$-agents tends to elevate system performance, it is essential to balance this with constraints such as budget, crowdedness, and diminishing utility gains, alongside ensuring effective control of $c$-agents.

The formal definition of the composition design as an optimization problem is formulated as follows: 
\begin{align*}
    P^{CD}: & \max_{N_c, \pi_c} \mathcal{F}(N_c, \pi_c) \\
          & s.t. V^{i_u,(\pi_u^{i_u}, \pi_u^{-i_u}, \pi_c)}(s) \geq V^{i_u,({\pi_u^{i_u}}', \pi_u^{-i_u}, \pi_c)}(s),  \\
          & \quad \forall i_u \in \mathcal{N}_u, \forall i_c \in \mathcal{N}_c, \forall s \in \mathcal{S}, \forall \pi_u^{i_u}, {\pi_u^{i_u}}' \in \Pi_u^{i_u},  \\
          & \quad \forall \pi_u^{-i_u} \in \Pi_u^{-i_u}, \pi_c \in \Pi_c,
\end{align*}
where $-i_u=\mathcal{N}_u\backslash i_u$, and $V^{i_u,\pi}$ is the value function of joint policy $\pi$ for $u$-agent $i_u$. Constraints set the conditions to be a Nash equilibrium (NE) policy for $u$-agents under the system parameter $N_c$ and $\pi_c$.

Addressing this problem is computationally challenging due to the need to determine optimal controllable agents, cooperative policies, and response strategies. These challenges arise primarily from the iterative process of finding equilibria and modifying system configurations, which must be repeated for each setup. Additionally, the lack of efficient mechanisms for transferring information across configurations exacerbates computational demands. This makes na\"ive BO-MARL approaches impractical and underscores the need for alternative methods to enable effective intervention.

\subsection{Hypernetworks}

Hypernetworks are neural networks designed to dynamically generate the weights of other networks, referred to as target networks. In traditional deep neural networks (DNNs), task-specific weights $\Theta_j$ are directly learned through backpropagation for each task $j$, using datasets $\{D_j = \{X_j, Y_j\}\}_{j=1}^{J}$, where $X_j$ and $Y_j$ represent input-output pairs. The mapping for task $j$ is expressed as $Y_j = \mathcal{G}(X_j; \Theta_j)$, with $\Theta_j$ being the learnable parameters. Hypernetworks offer an alternative approach by dynamically generating these weights using another network, denoted as $\mathcal{H}(C; \Phi)$, where $C$ is a task-specific context vector and $\Phi$ represents the learnable parameters of the hypernetwork. The generated weights are given by $\Theta_j = \mathcal{H}(c_j; \Phi)$. Unlike standard DNNs, where multiple sets of weights $\{\Theta_j\}_{j=1}^J$ are optimized independently, hypernetworks optimize only the shared parameters $\Phi$. This framework enables efficient learning across tasks by solving the optimization problem: 
\[
\mathrm{Hypernetwork}: \min_{\Phi} \mathcal{L}(\mathcal{G}(X; \mathcal{H}(C; \Phi)), Y),
\]
in contrast to a standard DNN, which minimizes:
\[
\mathrm{DNN}: \min_{\Theta_j} \mathcal{L}(\mathcal{G}(X_j; \Theta_j), Y_j), \forall j \in \{1, \cdots, J\},
\]
where $\mathcal{L}$ represents the loss function. 

\subsection{Mean-field reinforcement learning (MFRL)}

In MFRL, the complexity of multi-agent systems is addressed by approximating agent interactions through a mean-field approach \citep{yang2018mean}. Instead of modeling the influence of each individual agent, MFRL simplifies the actions of other agents as a mean action, capturing their average effect. Specifically, the action-value function $Q^i(s^i, a^i, a^{-i})$ for agent $i$ is approximated as $Q^i(s^i, a^i, a^{-i}) \approx \frac{1}{N(i)}\sum_{k \in N(i)} Q^i(s^i, a^i, a^k)$, where $N(i)$ denotes the neighboring agents of agent $i$. By defining a mean action $\bar{a}^i = \frac{1}{|N(i)|}\sum_{k \in N(i)} a^k$, the Q-function is further simplified to $Q^i(s^i, a^i, \bar{a}^i)$. The Boltzmann policy determines the probability of an agent selecting an action as $\pi_t^i(a^i | s, \bar{a}^i) = \frac{\exp(-\beta Q_t^i(s, a^i, \bar{a}^i))}{\sum_{a^{i'}} \exp(-\beta Q_t^i(s, a^{i'}, \bar{a}^i))}$, where $\beta$ is a temperature parameter. This approach enables each agent to optimize its actions based on the average field, thereby reducing computational complexity. However, MFRL relies on historical mean actions ($\bar{a}_{t-1}^i$) to compute current actions ($\pi_t^i(a^i | s, \bar{a}_{t-1}^i)$), potentially causing delays in agent interactions.

\section{Method}

\subsection{Hypernetwork-driven composition design}

In this study, we propose a novel framework that integrates a hypernetwork architecture to address the composition design problem in PCMAS. To the best of our knowledge, this is the first framework to leverage hypernetworks for this problem, effectively overcoming key challenges faced by conventional approaches such as BO-MARL. Unlike existing methods, which suffer from MARL bottlenecks and struggle with unexplored system configurations, the proposed method leverages hypernetworks to generalize across varying configurations.

The key innovation of this framework lies in its unified training regime, which efficiently generates policies for both $u$-agents and $c$-agents, facilitating effective information sharing across tasks. By employing hypernetworks, the computational cost associated with evaluating system configurations at the upper level is significantly reduced.

\begin{figure}[t]
    \centering
    \includegraphics[width=0.30\linewidth]{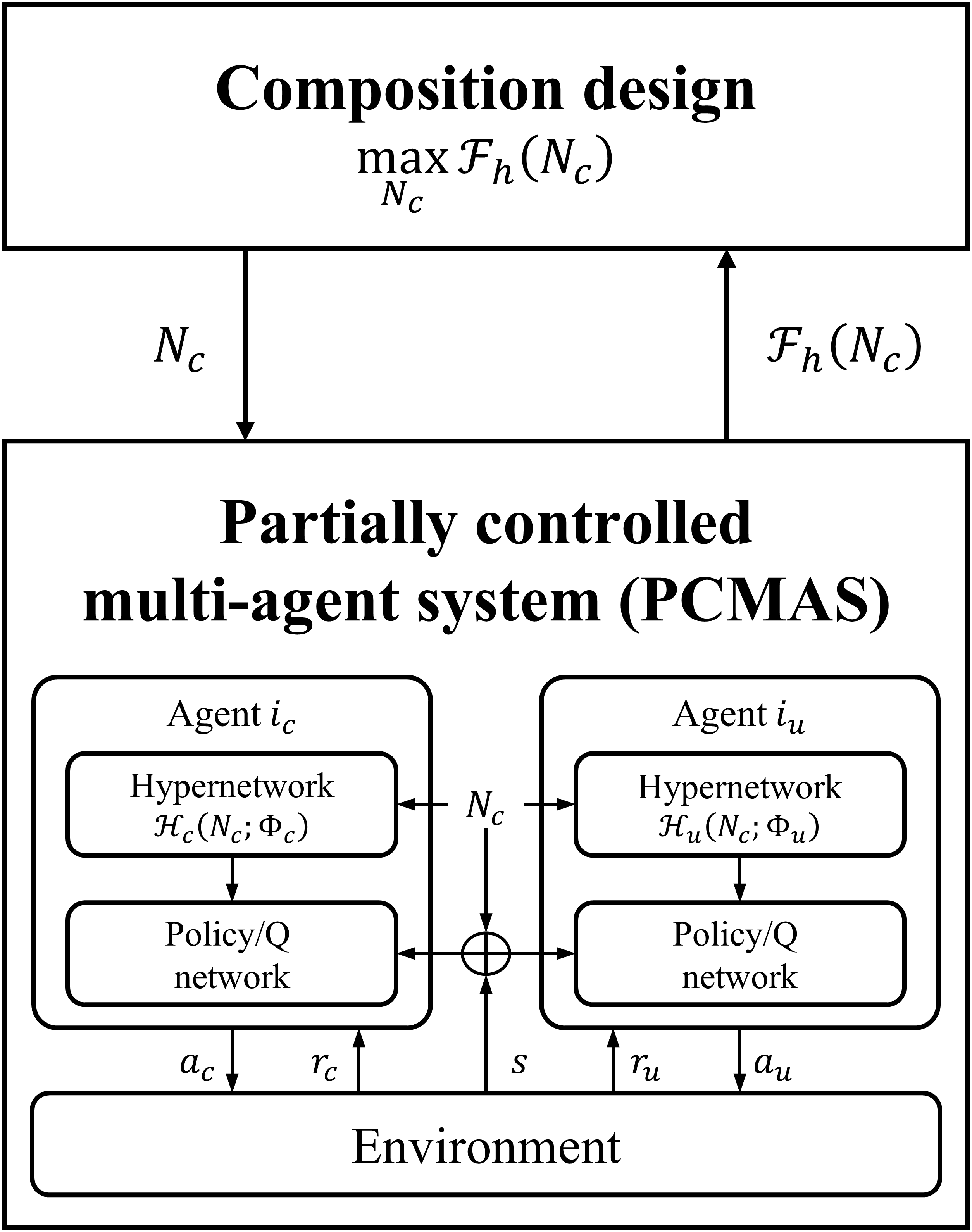}
    \caption{Hypernetwork-based architecture for solving the composition design problem. Policies for controllable ($c$-agents) and uncontrollable ($u$-agents) agents are generated by hypernetworks based on $N_c$. The system composition and environmental state are concatenated (denoted by $\oplus$) and used as input to the target network.}
    \label{fig:framework_simple}
\end{figure}

Figure~\ref{fig:framework_simple} provides an overview of the proposed framework. Specifically, given a designer's decision for $N_c$, the hypernetwork generates policies for $u$-agents and $c$-agents at the lower level, thereby addressing the inherent complexity of PCMAS. The objective function $\mathcal{F}(N_c, \pi_c)$ is approximated as a function of $N_c$ using the hypernetwork as:
\[
    \mathcal{F}(N_c, \pi_c) \approx \mathcal{F}_h(N_c).
\]
The PCMAS framework comprises two types of agents: $c$-agents ($i_c$) and $u$-agents ($i_u$). Each agent type employs a two-layered architecture. Specifically, hypernetworks $\mathcal{H}_c(N_c; \Phi_c)$ for $c$-agents and $\mathcal{H}_u(N_c; \Phi_u)$ for $u$-agents generate parameters for their respective target policy/Q networks. These networks make decisions or evaluate action quality based on the concatenation of the environment state $s$ and system configuration $N_c$, represented by the $\oplus$ symbol in Figure~\ref{fig:framework_simple}. The resulting actions, $a_u$ and $a_c$, are executed in the environment, which subsequently returns rewards ($r_u$ and $r_c$) and the next state. The objective values $\mathcal{F}_h(N_c)$ are derived by running the generated policies in simulation.

To optimize system composition, it is necessary to evaluate $\mathcal{F}_h(N_c)$ over multiple configurations. For a set of configurations $\{N_c^j\}_{j=1}^J$, identifying optimal policies $\{\pi_j(s;\Theta_j)\}_{j=1}^J$ is required (for clarity, policy notation is simplified here). Conventional methods require executing MARL multiple times per configuration to learn these policies, leading to significant computational overhead. In contrast, our framework efficiently generates policies via hypernetworks as $\{\pi_j(s;\mathcal{H}(N_c^j; \Phi))\}_{j=1}^J$, thereby reducing complexity in solving the upper-level optimization problem. This efficiency enables straightforward grid search to identify the optimal configuration $N_c^*$ or allows BO to minimize simulation runs further, underscoring the scalability and practicality of our approach.

\subsection{Reward parameter optimization}

In PCMAS, designing reward parameters for $c$-agents is crucial for system-level performance and achieving the designer's objectives. To address this, we propose a framework where reward parameter optimization, along with system configuration, is handled at the upper level. This allows a hypernetwork-based approach to jointly optimize reward parameters at the upper level and agents' policies at the lower level, offering a unified solution to the composition design problem.

In this framework, reward parameters of $c$-agents are no longer fixed or manually set; instead, they are treated as upper-level design variables. We assume the existence of a parameterized, abstract representation of the reward function. While directly optimizing the full reward function $r(s, a, s')$ is impractical, this abstraction enables optimization of reward parameters $\alpha_c$ as part of the upper-level objective. The hypernetwork at the lower level generates policies for both $u$-agents and $c$-agents, while upper-level optimization determines the optimal system configuration $N_c^*$ and reward parameters $\alpha_c^*$ that maximize the designer’s objective function $\mathcal{F}_h(N_c, \alpha_c)$.

\subsection{Mean action network}

To further enhance the scalability of the proposed framework, we incorporate a mean action network to improve predictions of other agents' mean behaviors. The network is trained via supervised learning approach, using true mean actions as labels. Formally, the network, denoted as $\mathcal{M}^i(\mathcal{S}, \mathcal{A}^i, N_c, \alpha_c;\Psi)$, is optimized to minimize the error between the predicted mean action $\hat{\bar{a}}_t^i$ and the true mean action $\bar{a}_t^i$. The loss function $\mathcal{L}_{\text{mean}}$ is defined as: 
$\mathcal{L}_{\text{mean}} = \frac{1}{n} \sum_{i=1}^{n} \|\bar{a}_t^i - \hat{\bar{a}}_t^i\|^2,$ where $n$ represents the number of training samples. Figure~\ref{fig:framework_all} illustrates the mean action network's structure, including its input and output. The network takes as input the current state $\mathcal{S}$, individual agent actions $\mathcal{A}^i$, and upper-level parameters $N_c$ and $\alpha_c$. It outputs an approximation of the mean action, $\hat{\bar{a}}^i$, which is used to enhance agent coordination and policy execution within the framework.

The action-value function $Q^i(s^i, a^i, a^{-i})$ for agent $i$ is then efficiently approximated as $Q^i(s^i, a^i, \hat{\bar{a}}^i)$. Consequently, the policy is then updated to $\pi_t^i(a^i | s, \hat{\bar{a}}_t^i)$ instead of relying on historical mean actions, i.e., $\pi_t^i(a^i | s, \bar{a}_{t-1}^i)$.

\subsection{Overall architecture}

\begin{figure}[t]
    \centering
    \includegraphics[width=0.40\linewidth]{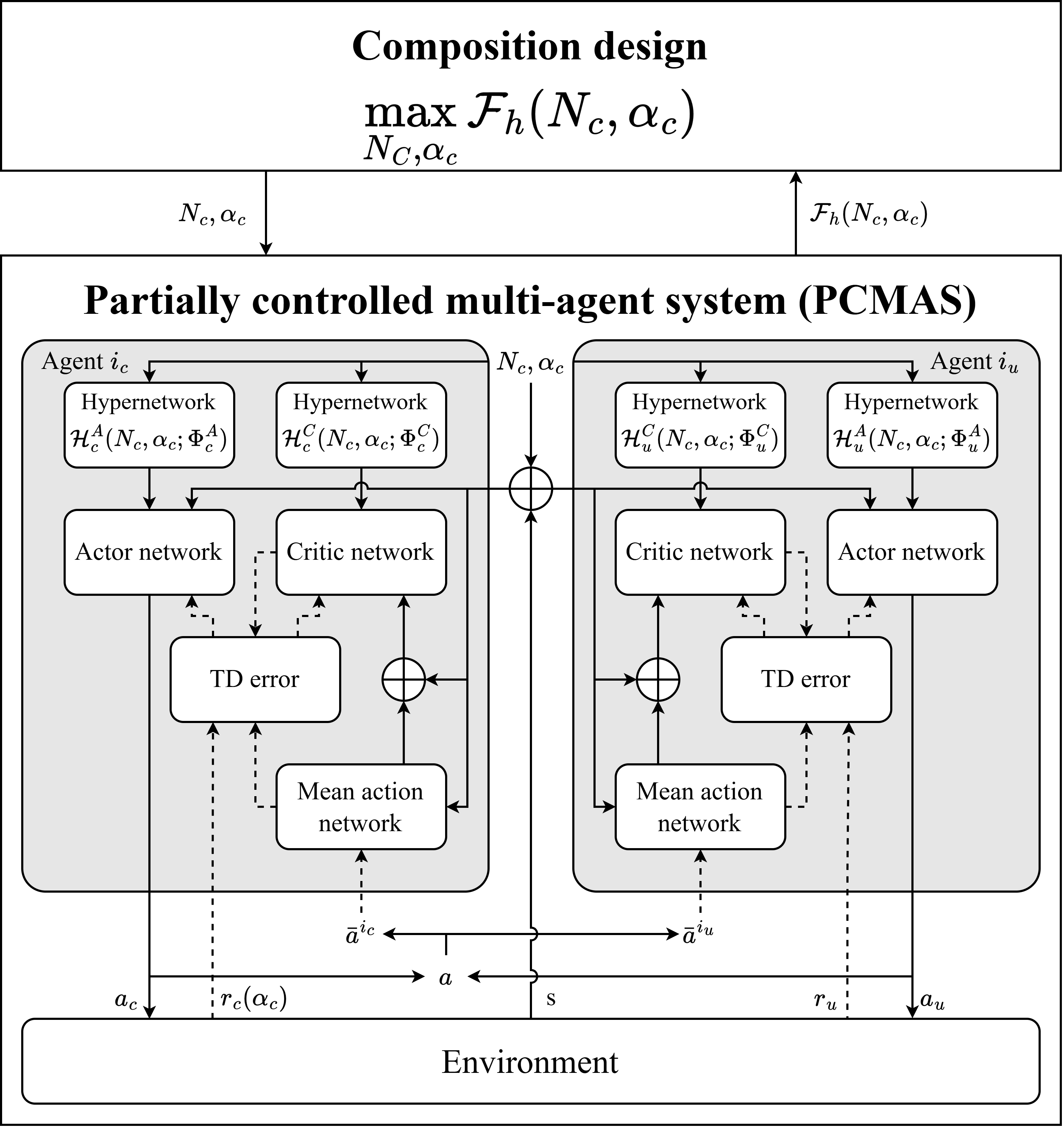}
    \caption{The overall architecture of the proposed framework. The dotted lines represent the network update process.}
    \label{fig:framework_all}
\end{figure}

Figure~\ref{fig:framework_all} illustrates the architecture of the proposed framework for composition design in PCMAS. The framework adopts a hierarchical structure that decouples the optimization of upper-level design variables, such as system configuration $N_c$ and reward parameters $\alpha_c$, from the generation of agent-specific policies. At the upper level, the design variables are optimized to maximize the designer's objective function $\mathcal{F}_h(N_c, \alpha_c)$. These variables are then provided as inputs to the hypernetwork at the lower level, which generates policies for $c$-agents and $u$-agents using a mean field actor-critic (MFAC) approach. This hierarchical structure ensures that global optimization is handled at the upper level, while policy generation is efficiently managed at the lower level.

\section{Experiments}

In this section, we evaluate the proposed method's effectiveness in balancing supply and demand for E-hailing driver repositioning using large-scale NYC taxi data. The experiments assess the framework's performance in several aspects: (1) demonstrating its superiority over existing methods in approximating equilibrium policies, highlighting the utility of hypernetworks for modeling lower-level behaviors; (2) optimizing system performance under various scenarios; (3) providing insights into the efficient use of controllable agents; and (4) analyzing the impact of the mean action network through ablation studies. Additional experiments comparing hypernetwork training across the entire design space versus segmented training are presented in Appendix~\ref{app_subsec:exp_segment}.

\subsection{Driver repositioning}

\subsubsection{Environment}

E-hailing driver repositioning represents an environment where taxi drivers aim to maximize their individual rewards by selecting profitable passengers, while the platform seeks to optimize overall system performance. System performance is measured using metrics such as the order response rate (ORR) \citep{shou2020reward,xie2023two}. To improve driver distribution, the platform employs a composition design approach, contracting with a portion of uncontrollable agents at a cost to align their actions with system objectives. However, to avoid excessive operational costs, it is crucial for the platform to maintain a reasonable profitability ratio (PR). The system's objective is formulated as follows:
\begin{align}
    \mathcal{F} = k \times \text{ORR} + (1-k) \times \text{PR},
    \label{eq:objective_function}
\end{align}
where \(k \in [0,1]\) represents the trade-off parameter between ORR and PR. In Equation~(\ref{eq:objective_function}), ORR is defined as the proportion of served requests relative to the total number of requests. PR, on the other hand, is calculated as the net revenue from served fares after deducting hiring costs, normalized by total fares. A detailed description of this environment is provided in Appendix~\ref{app_subsec:environment}.

\begin{figure}[t]
    \centering
    \includegraphics[width=0.40\linewidth]{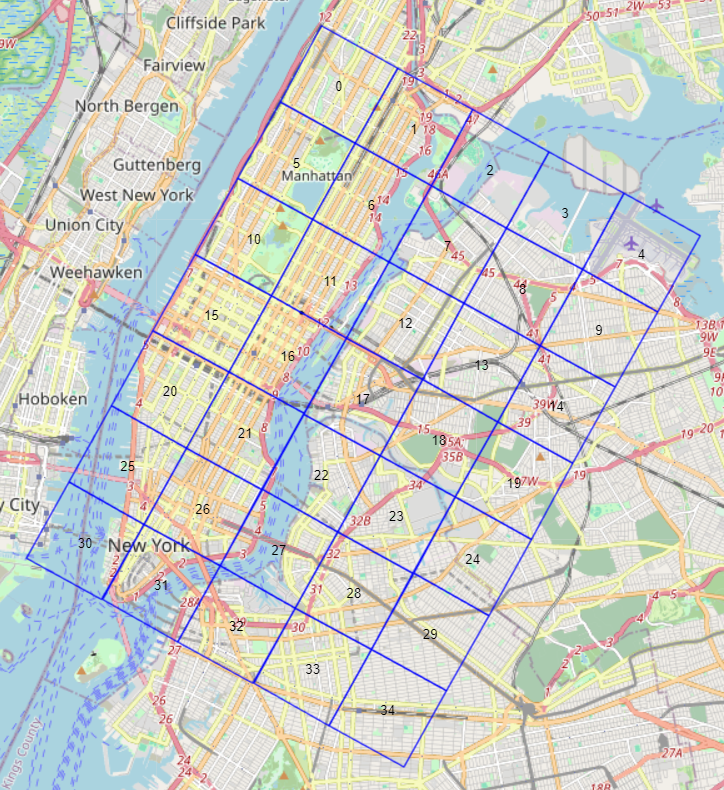}
    \caption{Geographical representation of the study area, covering zones from Manhattan to LaGuardia Airport.}
    \label{fig:experiment_NYC_map}
\end{figure}

To evaluate the proposed model, we use a real-world large-scale taxi dataset from the New York City (NYC) Taxi \& Limousine Commission\footnote{\url{https://www1.nyc.gov/site/tlc/about/tlc-trip-record-data.page}}. Specifically, we analyze weekday data for yellow and green taxis in May 2014. The study focuses on zones from Manhattan to LaGuardia Airport, represented as a 7$\times$5 grid with a 2 km resolution, as shown in Figure~\ref{fig:experiment_NYC_map}. The time interval of interest is restricted to evening peak hours, i.e., 4 PM to 8 PM, comprising 21 time steps at 12-minute intervals. Additional dataset details are in Appendix~\ref{app_subsec:data}.

\subsubsection{Reward function of controllable agents}

In the driver repositioning problem, uncontrollable agents receive a reward $r_u = \text{fare}$, where $\text{fare}$ represents the actual fare associated with a request. Conversely, to enhance system-level performance, we define a synthetic reward function for controllable agents as follows: $r_c(\alpha) = c \cdot (1 - SC_l(\alpha))$, where $c$ denotes the constant synthetic fare assigned to controllable agents, and $SC_l(\alpha)$ is the service charge ratio for grid $l$. This formulation is based on the assumption that indistinguishable requests can improve the overall ORR. To further optimize the system objective $\mathcal{F}$, the service charge is imposed exclusively on oversupplied grids and is proportional to their congestion levels. The congestion level is quantified by the demand-to-supply ratio ($DS_l$), where a lower $DS_l$ indicates an excess supply of drivers relative to demand. The service charge ratio $SC_l$ incorporates an adjustable parameter $\alpha$, referred to as the penalty strength, and is defined as follows:
\[
SC_l(\alpha) =
\begin{cases}
    \alpha \cdot (1 - DS_l), & \text{if } DS_l \leq 1, \\
    0, & \text{otherwise.}
\end{cases}
\]

Accordingly, the resulting expression for the reward function for controllable agents can be written as:
\[
r_c(\alpha) = c \cdot (1 - \alpha \cdot (1 - DS_l) \cdot \mathds{1}_{DS_l \leq 1}),
\]
where $\mathds{1}_{DS_l \leq 1}$ is an indicator function that equals 1 if $DS_l \leq 1$, and 0 otherwise. In this experiment, $\alpha$ takes values in the range $[0, 1]$, and we aim to find the optimal $\alpha$ that maximizes the system objective using a hypernetwork.

\subsubsection{Baselines}

In our proposed framework, the target network, which is generated by the hypernetwork, incorporates the MFAC algorithm. Specifically, we implement MFAC using a MLP architecture. Importantly, this implementation is flexible and can be replaced with any standard deep RL method. To demonstrate that our framework approximates NE policies more effectively than other algorithms, we ensure that the baseline algorithms (non-hypernetwork-based) adopt the same architecture as our target network. The baselines employed in our experiments are as follows: (1) Target: a standard deep RL method with the same structure as the target network, taking only the state as input. (2) AugTarget: extends Target by including upper-level parameters (i.e., controllable agents, reward parameters) in its input. (3) Target-Large: increases neurons per layer to match our framework's total learnable parameters. (4) AugTarget-Large: similar to Target-large, but with augmented inputs as in AugTarget. Further details on baselines and hyperparameters are provided in Appendix~\ref{app_subsec:baselines} and Appendix~\ref{app_subsec:hyperparameters}, respectively.

\subsubsection{Approximate NashConv}

The NashConv metric is employed to evaluate the deviation of a given policy $\pi$ from the NE policy \citep{li2023population}. Formally, for the policy $\pi$, the NashConv value is defined as: 
\[
\text{NashConv}(\pi) = \sum_{i=1}^N \max_{\hat{\pi}^i} V^i(s^i, \hat{\pi}^i, \pi^{-i}) - V^i(s^i, \pi),
\]
where $N$ denotes the number of agents, $V^i$ represents the value function for agent $i$, $\hat{\pi}^i$ is the best response (BR) policy for agent $i$, and $\pi^{-i}$ denotes the policies of all agents except agent $i$. This metric quantifies how far the policy $\pi$ is from satisfying the NE conditions. Specifically, if $\text{NashConv}(\pi) = 0$, then $\pi$ corresponds to a NE policy.

In practice, computing the exact NashConv is often infeasible in complex games due to the difficulty of determining exact BR policies. To address this challenge, we approximate the BR policy for a representative agent by training a new policy while other agents follow the current policy. This training process is repeated for each composition and reward parameter configuration of the game.

The approximated NashConv value is then calculated as the difference between the representative agent's reward when following its BR policy and its reward when following the generated policy. This approach provides an estimate of how closely the generated policy aligns with NE. Additional details on this computation are provided in Appendix~\ref{app_subsec:environment_nashconv}.

\subsection{Results}

\subsubsection{NashConv comparison for policy evaluation}\label{subsubsec:experiments_nashconv}

\begin{figure}[t]
    \centering
    \includegraphics[width=0.50\linewidth]{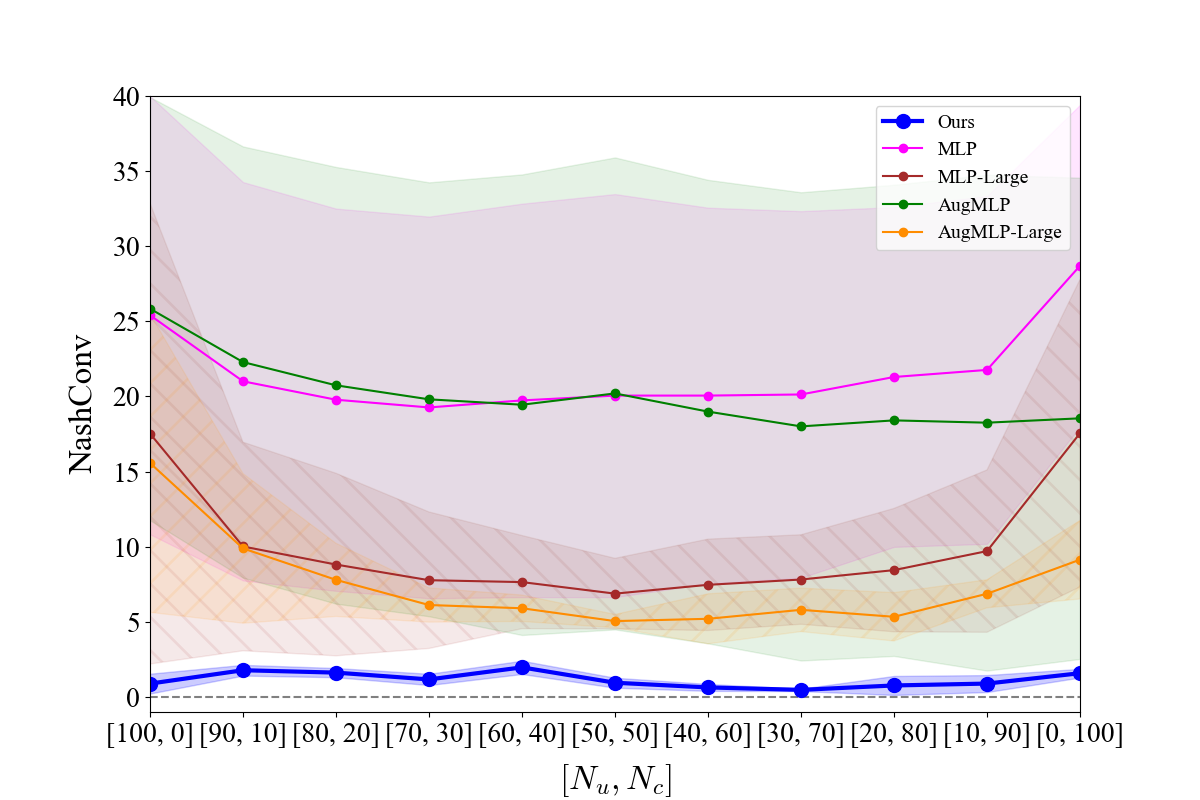}
    \caption{Comparison of NashConv values across our algorithm and baselines when $\alpha=0.00$.}
    \label{fig:result_nashconv_alpha=0.00}
\end{figure}

We conducted experiments with a total of 100 agents. Hypernetworks were trained across all possible compositions and reward parameters. Subsequently, BR policies were trained for each configuration, which included both composition and reward parameters, at regular intervals for a representative agent. For each agent type, NashConv was computed as the difference between the served fare of the representative agent and the average served fare of agents belonging to that specific type. The experiments were repeated using three random seeds, and 100 simulations were performed for each configuration to obtain the corresponding objective values. The results, shown in Figure~\ref{fig:result_nashconv_alpha=0.00}, demonstrate that our approach significantly outperforms existing methods in approximating equilibrium policies. The figure plots mean values across seeds with shaded areas indicating standard deviation. Notably, our algorithm achieves NashConv values near zero across all compositions, indicating that the generated policies closely approximate NE policies and eliminate the need to train separate networks for each configuration. This underscores the potential of hypernetworks in modeling lower-level agent behaviors. Additional experiments with different reward parameters yielded consistent results, as detailed in Appendix~\ref{app_subsec:exp_nashconv}.

\subsubsection{System optimization under various scenarios}\label{subsubsec:experiments_system_optimization}

\begin{figure}[t]
    \centering
    \includegraphics[width=0.5\linewidth]{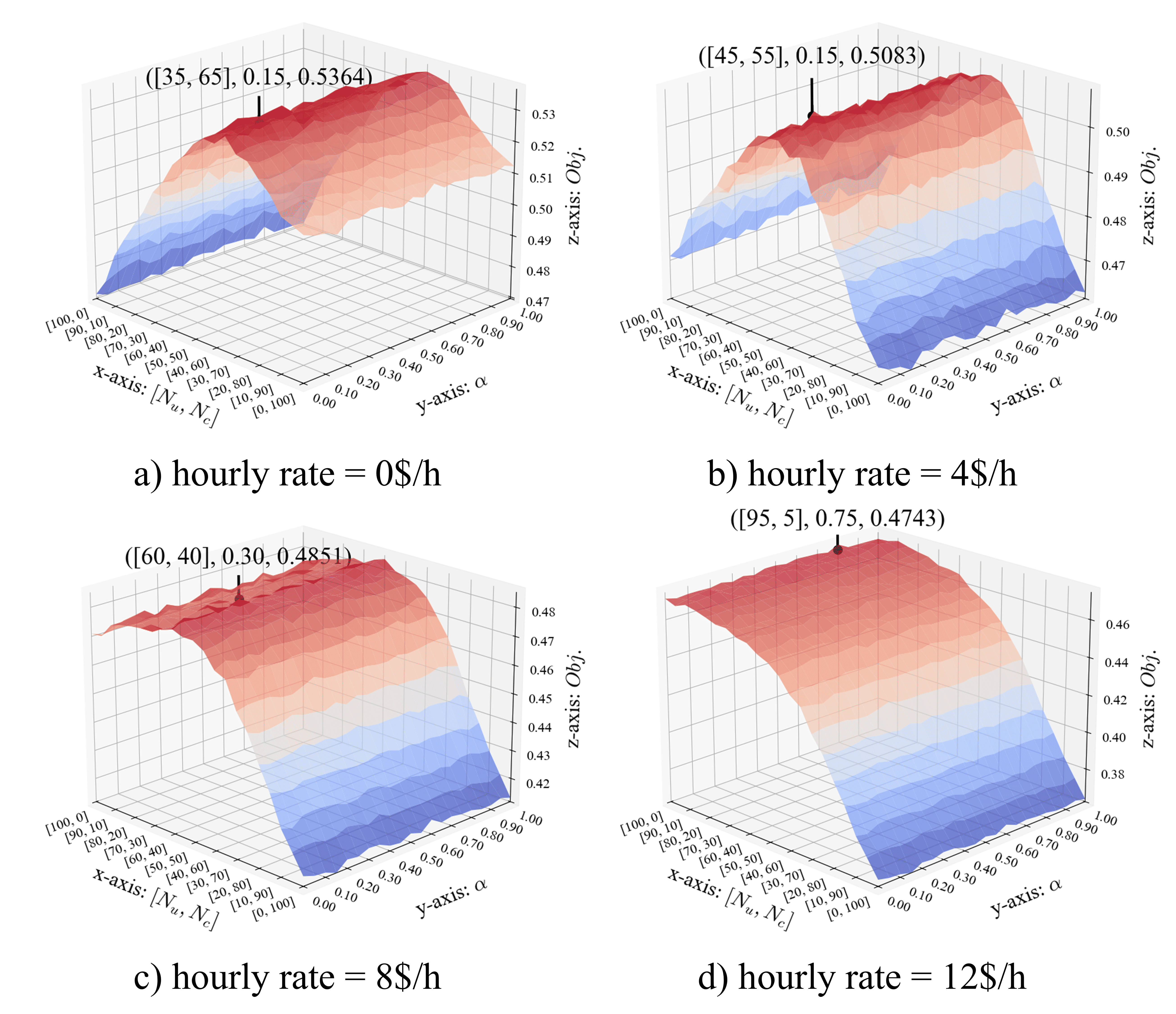}
    \caption{Objective value improvements for varying hourly rates with $k=0.6$.}
    \label{fig:result_optimization_k=0.6}
\end{figure}

After validating the hypernetwork through the NashConv experiment, experiments were conducted to demonstrate that the proposed framework effectively adapts and optimizes system performance under various scenarios. Specifically, these experiments identified the optimal composition and reward parameters for various system configurations. We optimized the objective function in Equation~(\ref{eq:objective_function}), with adjustments to the trade-off parameter $k$ and the hourly rate for hiring agents. Policies were generated for each configuration, and the average objective value was calculated over 100 simulation runs. The results for different hourly rates when $k=0.6$ are presented in Figure~\ref{fig:result_optimization_k=0.6}. Without intervention, the baseline objective value was 0.4710. By contrast, the proposed framework improved the objective value ranging from 0.70\% to 13.89\%, depending on the configuration. For instance, when the hourly rate was set to 4\$/h, employing 55 agents and using policies generated with a reward parameter of 0.15 led to a 7.29\% improvement in system performance. Additional detailed results are provided in Appendix~\ref{app_subsec:exp_system_optimization}. % 13.89\%, 7.92\%, 2.99\%, and 0.70\%. 

These results demonstrate that the proposed framework effectively enhances system performance metrics by optimizing composition design within PCMAS. The hypernetwork-based approach facilitates efficient evaluation of the design space, enabling the identification of optimal configurations across diverse scenarios. 
This represents a significant advantage over traditional methods such as BO-MARL, which are limited to specific objectives and struggle to adapt to varying conditions. Notably, system performance was highest when controllable and uncontrollable agents behaved complementarily, emphasizing the importance of synergy in agent interactions.

\subsubsection{Utility of controllable agents}

\begin{figure}[t]
    \centering
    \includegraphics[width=0.5\linewidth]{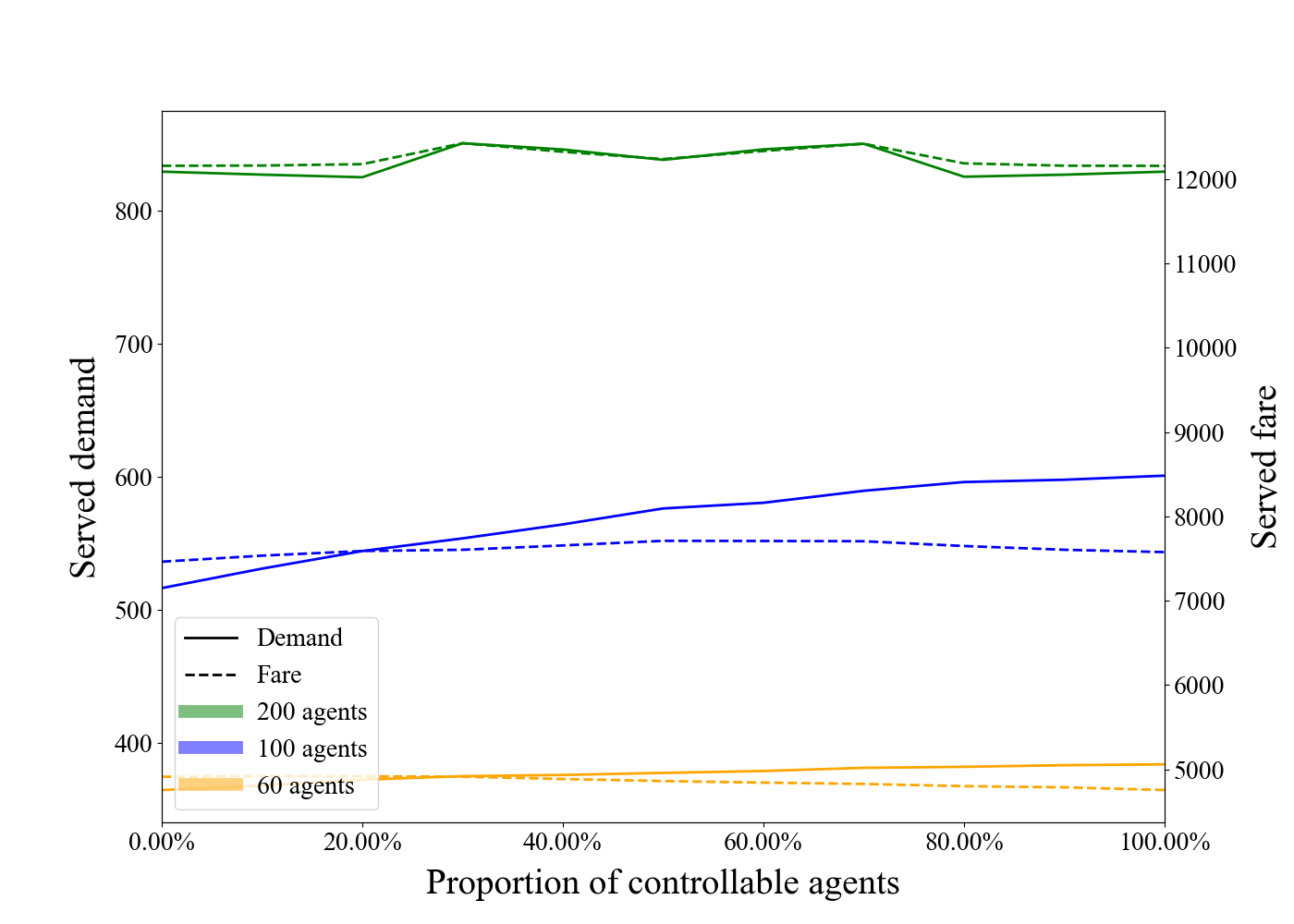}
    \caption{Utility of controlling agents on served demands and fares across different total number of agents.}
    \label{fig:result_utility_of_controlling_agents}
\end{figure}

The experiment aimed to assess the effectiveness of controllable agents in improving system performance, focusing on their impact on served demands and fares. Various configurations were tested by altering both the total number of agents (60, 100, and 200), the proportion of controllable agents, and reward parameters. For each configuration, 100 simulations were performed to obtain the corresponding objective values. To highlight the effect of agent numbers, the results were averaged over reward parameters and are presented in Figure~\ref{fig:result_utility_of_controlling_agents}.

When the total number of agents was set to 100, increasing the proportion of controllable agents significantly improved served demands by 16.37\%, while served fares increased by 1.53\%, remaining relatively stable regardless of agent composition. In contrast, with 200 agents, neither served demands nor fares showed any noticeable changes. For scenarios with 60 agents, introducing controllable agents moderately increased served demands by 5.29\% but reduced served fares by 3.20\%.

These results suggest that controllable agents are most beneficial within a specific range of system size. When the number of agents is too low relative to demand, there is sufficient demand for all agents regardless of their controllability, leading to minimal performance differences between controllable and uncontrollable agents. Conversely, when the number of agents is too high relative to demand, even uncontrollable agents acting selfishly can still serve most demands. 

\subsubsection{Ablation on the mean action network}

\begin{figure}[t]
    \centering
    \includegraphics[width=0.50\linewidth]{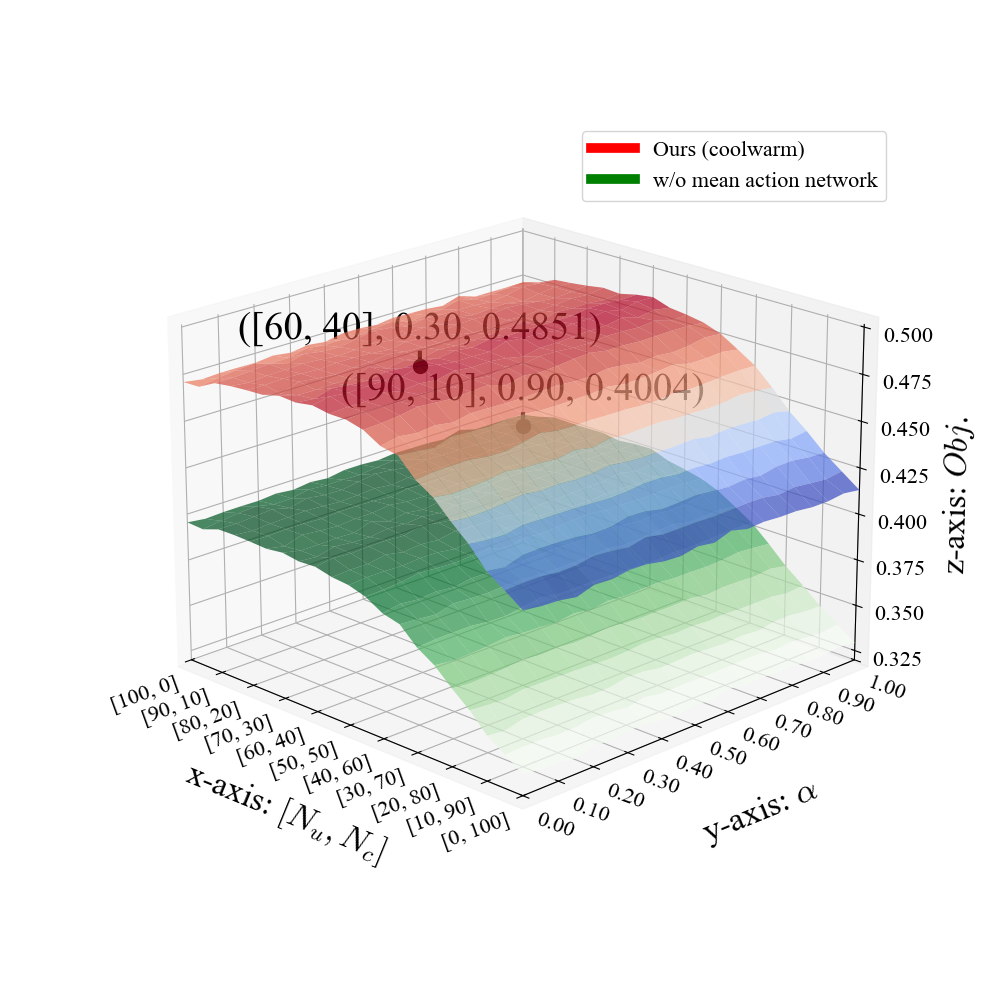}
    \caption{Impact of the mean action network on performance}
    \label{fig:result_ablation}
\end{figure}

The ablation study on the mean action network highlights its essential role in improving the performance of the proposed framework. Specifically, the experiment compared two scenarios: one where agents utilized the predicted mean action obtained through the mean action network for decision-making ($\pi_t^i(a^i | s, \hat{\bar{a}}_t^i)$), and another where agents relied on the mean action from the previous time step ($\pi_t^i(a^i | s, \bar{a}_{t-1}^i)$). The results, averaged over 100 simulation runs, revealed that removing the mean action network led to a decrease in the objective value by 17.46\%, as illustrated in Figure~\ref{fig:result_ablation}. This finding emphasizes the effectiveness of the network in improving system coordination and scalability. Furthermore, the observed performance degradation during the ablation study underscores the importance of real-time approximation of mean actions ($\hat{\bar{a}}_t^i$) rather than depending on historical values. These findings confirm that integrating the mean action network not only improves real-time decision making, but also significantly contributes to achieving scalability by reducing computational overhead and improving coordination efficiency among agents.

\section{Conclusion}

In this study, we proposed a novel hypernetwork-based framework to address the composition design problem in PCMAS. Our approach integrates hypernetworks to efficiently generate policies for both controllable and uncontrollable agents, enabling significant computational savings and enhanced information sharing across system configurations. By incorporating reward parameter optimization and a mean action network, our framework further improves scalability and policy performance in large-scale systems.

The experimental results, conducted using real-world NYC taxi data, demonstrated the framework's ability in achieving near-equilibrium policies and optimizing system performance under diverse conditions. Specifically, our method outperformed existing approaches in approximating equilibrium policies, while effectively balancing computational complexity. After validating our framework, we demonstrated that an objective function incorporating the order response rate (ORR) and the profitability ratio (PR) can be improved by up to 13.89\%. Moreover, our analysis showed that the improvements by deploying controllable agents are most pronounced within specific ranges of system size, offering valuable insights into the effective utilization of controllable agents.

Overall, this work highlights the potential of hypernetworks to transform system-level optimization in PCMAS by providing a unified and scalable approach to policy generation and composition design.

% \bibliographystyle{abbrvnat}
% \bibliography{neurips_2024}

%%%%%%%%%%%%%%%%%%%%%%%%%%%%%%%%%%%%%%%%%%%%%%%%%%%%%%%%%%%%

\newpage

\appendix

\section{Appendix / supplemental material}

\begin{algorithm}[t!]
    \caption{Pseudo-code of the hypernetwork-based composition design framework}
    \label{alg:CD_HCD}
    \begin{flushleft}
        \textbf{Inputs:} Hyperparameters
    \end{flushleft}
    \begin{algorithmic}[1]
        \STATE Initialize hypernetworks $\mathcal{H}_c$ and $\mathcal{H}_u$ following the initialization method of \citep{sarafian2021recomposing} and mean action networks $\mathcal{M}_c$ and $\mathcal{M}_u$
        \STATE Set replay buffer $\mathcal{B} \gets \emptyset$
        \FOR{$episodes=1,2,\cdots,$}
            \STATE Sample $N_c$ and $\alpha_c$ from an uniform distribution and build a game $G(N_c, \alpha_c)$
            \STATE Generate policies $\pi_c^{i_c}=\mathcal{H}_c(N_c, \alpha_c;\Phi_c)$ and $\pi_u^{i_u}=\mathcal{H}_u(N_c, \alpha_c;\Phi_u)$
            \STATE Set $t \gets 0$
            \WHILE{$t<T$}
                \STATE Get predicted mean actions $\hat{\bar{a}}_t^{i_c}\sim \mathcal{M}_c, \forall i_c \in \mathcal{N}_c$ and $\hat{\bar{a}}_t^{i_u}\sim \mathcal{M}_u, \forall i_u \in \mathcal{N}_u$
                \STATE Sample actions $a_t^{i_c}\sim\pi_c^{i_c}(\cdot|s_t^{i_c}, \hat{\bar{a}}_t^{i_c}), \forall i_c \in \mathcal{N}_c$ and $a_t^{i_u}\sim\pi_u^{i_u}(\cdot|s_t^{i_u}, \hat{\bar{a}}_t^{i_u}), \forall i_u \in \mathcal{N}_u$
                \STATE Execute $a_t^i$ in $G(N_c, \alpha_c)$, $\forall i\in\mathcal{N}$
                \STATE Observe new state $s_{t+1}^i$, $\forall i\in\mathcal{N}$
                \STATE Receive rewards $r_t^i=r^i(s_t^i, a_t^i, s_{t+1}^i)$, $\forall i\in\mathcal{N}$
                \STATE Store data $\mathcal{B} \gets \mathcal{B} \cup \{(s_t, a_t, r_t, s_{t+1})\}$
                \STATE Set $t \gets t+1$
            \ENDWHILE
            \IF{$episode\%E=0$}
                \STATE Update hypernetworks $\mathcal{H}_c$ and $\mathcal{H}_u$
                \STATE Update mean action networks $\mathcal{M}_c$ and $\mathcal{M}_u$
            \ENDIF
        \ENDFOR
    \end{algorithmic}
    \begin{flushleft}
        \textbf{Return} Trained hypernetworks $\mathcal{H}_c$ and $\mathcal{H}_u$, and trained mean action networks $\mathcal{M}_c$ and $\mathcal{M}_u$
    \end{flushleft}
\end{algorithm}

The pseudo-code of the training procedure for our hypernetwork-based composition design framework is illustrated in Algorithm~\ref{alg:CD_HCD}. This procedure involves the training of hypernetworks through data derived from multiple games, embodying a multi-task training approach. Specifically, at the beginning of each episode, a game $G(N_c, \alpha_c)$ is constructed by uniformly sampling a value for $N_c$ and $\alpha_c$. Subsequently, the hypernetworks, upon receiving $N_c$ and $\alpha_c$ as input, generate respective policies. These policies are used by agents in their interactions within the environment for $T$ steps. The experiences gathered during these steps are stored as tuples within the replay buffer $\mathcal{B}$. Finally, at intervals of every $E$ episodes, hypernetworks are updated with the collected experience tuples. 

\section{Experiment}

\subsection{Environment}\label{app_subsec:environment}

The driver repositioning problem is modeled as a partially observable MG $<N, S, \{A^i\}, P, \{r^i\}, \{O^i\}, \gamma>$ \citep{hansen2004dynamic}, where multiple drivers have to decide which grid to travel in order to pick up a passenger and receive rewards based on the fare of each trip. Because the environmental state $s \in S$ is not fully observable, each agent draws an individual observation $o^i \in O^i$, where $O^i \triangleq \{o^i | s \in S, o^i=\Omega(s,i)\}$ is the set of observations for agent $i$, and $\Omega: S \times N \mapsto \mathbb{R}^q$ is the observation function. The observation of agent $i$ consists of its location $l$ and current time $t$. $A^i$ is the set of actions for agent $i$, where $a^i \in A^i$ can be any of the five possible actions, i.e., moving into any of its neighboring grids or staying in the current grid. Agent $i$ selects the action $a^i$ using the policy $\pi^i: O^i \times A^i \mapsto [0, 1]$. After drivers' movements, passenger requests and drivers in the same grid will be matched. Drivers can access individual reward $r^i$ by fulfilling a passenger request. When the driver picks up the passenger, it drives to the destination of the passenger. 

\subsection{Data}\label{app_subsec:data}

For our experiments, we use a real-world large-scale taxi dataset provided by the New York City (NYC) Taxi \& Limousine Commission. Specifically, we analyze weekday data in May 2014, focusing on yellow and green taxi trips. The study focuses on zones from Manhattan to LaGuardia Airport. The time interval of interest is restricted to evening peak hours, i.e., 4 PM to 8 PM. A representative sample of the dataset is presented in Table~\ref{tab:taxi_data_sample}. On average, the NYC taxi data records approximately 94,520 trips during evening peak hours each day, with 89,208 of these trips involving travel between Manhattan and LaGuardia Airport. We constructed the environment that reflects these demand distributions. For the experiments, we scaled down the demand volume by a factor of 1/60 to facilitate computational efficiency and practical testing.

\begin{table*}[t]
    \centering
    {\small
        \begin{tabular}{@{\extracolsep{\fill}}llllr}
            \toprule
            Pickup datetime     & Dropoff datetime    & Pickup coordinate       & Dropoff coordinate      & Fare \\
            \midrule
            2024-05-03 19:42:37 & 2024-05-03 19:48:14 & (-73.895073, 40.754677) & (-73.915863, 40.752468) & 5.5  \\
            2024-05-11 18:57:04 & 2024-05-03 19:23:34 & (-73.986313, 40.689182) & (-73.989334, 40.630630) & 24   \\
            \bottomrule
        \end{tabular}
    }
    \caption{Taxi data sample.}
    \label{tab:taxi_data_sample}
\end{table*}

% \begin{table*}[t]
%     \centering
%     \begin{tabular*}{\textwidth}{@{\extracolsep{\fill}}llllr}
%         \toprule
%         Pickup datetime     & Dropoff datetime    & Pickup coordinate       & Dropoff coordinate      & Fare \\
%         \midrule
%         2024-05-03 19:42:37 & 2024-05-03 19:48:14 & (-73.895073, 40.754677) & (-73.915863, 40.752468) & 5.5  \\
%         2024-05-11 18:57:04 & 2024-05-03 19:23:34 & (-73.986313, 40.689182) & (-73.989334, 40.630630) & 24   \\
%         \bottomrule
%     \end{tabular*}
%     \caption{Taxi data sample.}
%     \label{tab:taxi_data_sample}
% \end{table*}

\subsection{Hyperparameters}\label{app_subsec:hyperparameters}

The proposed framework adopts a hierarchical structure comprising two main components: the hypernetwork and the target network. The hypernetwork is responsible for generating the weights of the target network, while the target network serves as either the policy network (actor) or the Q-network (critic). The target networks for the actor and critic are implemented as multilayer perceptrons (MLPs) with three hidden layers. Specifically, the actor's target network has layers with (32, 16, 18) neurons, while the critic's target network consists of layers with (64, 32, 16) neurons. Similarly, the hypernetworks for both components are MLPs but with two hidden layers. The hypernetwork for the actor has hidden layers with (128, 64) neurons, whereas that of the critic has two layers with (128, 128) neurons. Additionally, the mean action network is structured as an MLP with three hidden layers containing (32, 16, 8) neurons.

The training process for the hypernetworks involves a total of 60,000 episodes. Different learning rates are applied to various components: 0.00004 for the actor, 0.0003 for the critic, and 0.0001 for the mean action network. Furthermore, the discount factor ($\gamma$) is set to a value of 1. This configuration reflects an assumption that drivers do not differentiate between fares collected within a single day, effectively treating all rewards as equally significant over this time horizon.

Furthermore, the initialization method proposed by \citep{sarafian2021recomposing} is employed, addressing the limitation of conventional DNN initialization techniques, such as those of Xavier \citep{glorot2010understanding} and Kaiming initialization \citep{he2015delving}, which do not guarantee uniform initialization ranges for the weights of the target network. When our experiments are conducted with Xavier \citep{glorot2010understanding} or Kaiming initialization \citep{he2015delving}, the experiments start with a very biased policy for most random seeds, resulting in poor learning.

\subsection{Baselines}\label{app_subsec:baselines}

As the network architecture of our proposed framework is large (has more learnable parameters), in addition to the standard baselines (MLP and AugMLP), we consider two more baselines: MLP-Large and AugMLP-Large, which have similar numbers of learnable parameters as our framework by increasing the number of neurons of the hidden layers of the network. This is critical to ensure a fair comparison and demonstrate the effectiveness of our approach. In Table~\ref{tab:number_of_learnable_parameters}, we provide the numbers of learnable parameters of all methods.

\begin{table}[t]
    \centering
    \begin{tabular}{lr}
        \toprule
        Algorithm    & Number of parameters \\
        \midrule
        Ours         & 499.9k               \\
        MLP          & 3.8k                 \\
        AugMLP       & 86.8k                \\
        MLP-Large    & 506.4k               \\
        AugMLP-Large & 479.0k               \\
        \bottomrule
    \end{tabular}
    \caption{The numbers of learnable parameters of different methods.}
    \label{tab:number_of_learnable_parameters}
\end{table}

\subsection{Approximate NashConv}\label{app_subsec:environment_nashconv}

To approximate NashConv, we train the BR policy of a representative agent, while keeping the policies of other agents fixed. In our experiments, we select a specific agent $i$ (which can be either a $u$-agent or a $c$-agent), along with a target composition and a reward parameter, and then train the BR policy for that agent. The composition space and reward parameter space are each divided into 10 equal segments. For instance, the tested compositions include $[[100, 0], [90, 10], \cdots, [10, 90], [0, 100]]$. 

The BR network is implemented using an actor-critic framework. Both the actor and critic are modeled as MLPs with three hidden layers. The actor network consists of layers with (64, 32, 16) neurons, while the critic network has layers with (128, 64, 32) neurons. The training process spans 60,000 episodes in total. Different learning rates are applied to the actor and critic networks: 0.00004 for the actor and 0.0003 for the critic. Additionally, the discount factor ($\gamma$) is set to 1 throughout the training process.

\section{Experimental results}\label{app_sec:experimental_results}

\subsection{NashConv comparison for policy evaluation}\label{app_subsec:exp_nashconv}

\begin{figure}[t]
    \centering
    \includegraphics[width=0.50\linewidth]{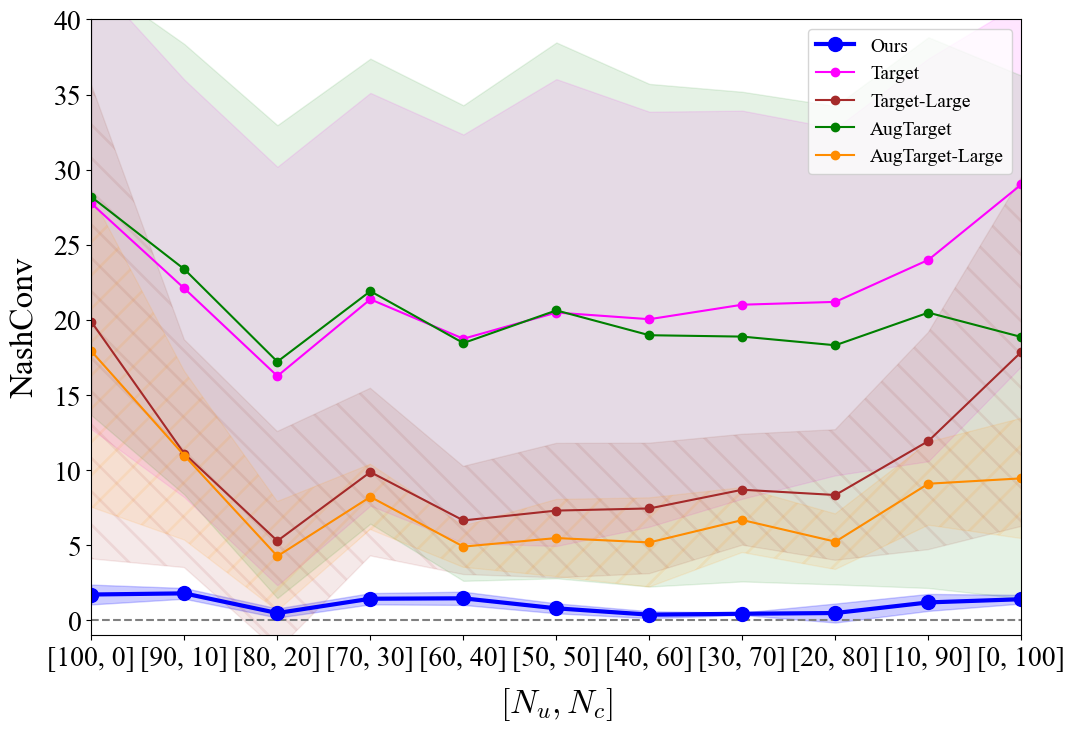}
    \caption{Comparison of NashConv values across our algorithm and baselines when $\alpha=0.50$.}
    \label{fig:result_nashconv_alpha=0.50}
\end{figure}

\begin{figure}[t]
    \centering
    \includegraphics[width=0.50\linewidth]{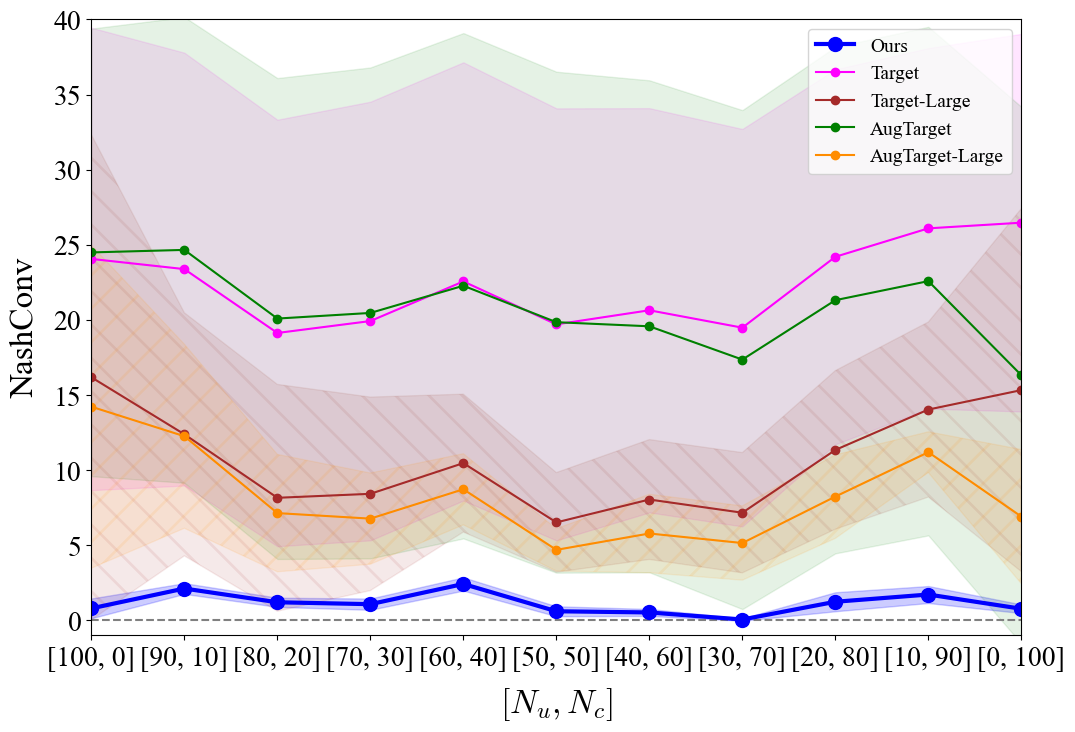}
    \caption{Comparison of NashConv values across our algorithm and baselines when $\alpha=1.00$.}
    \label{fig:result_nashconv_alpha=1.00}
\end{figure}

The NashConv results for different reward parameters ($\alpha=0.50, 1.00$) are presented in Figure~\ref{fig:result_nashconv_alpha=0.50} and \ref{fig:result_nashconv_alpha=1.00}. In all experiments, our algorithm achieves NashConv values near zero across all compositions and outperforms existing methods in approximating equilibrium policies. 

\subsection{System optimization under various scenarios}\label{app_subsec:exp_system_optimization}

\begin{figure}[t]
    \centering
    \includegraphics[width=0.50\linewidth]{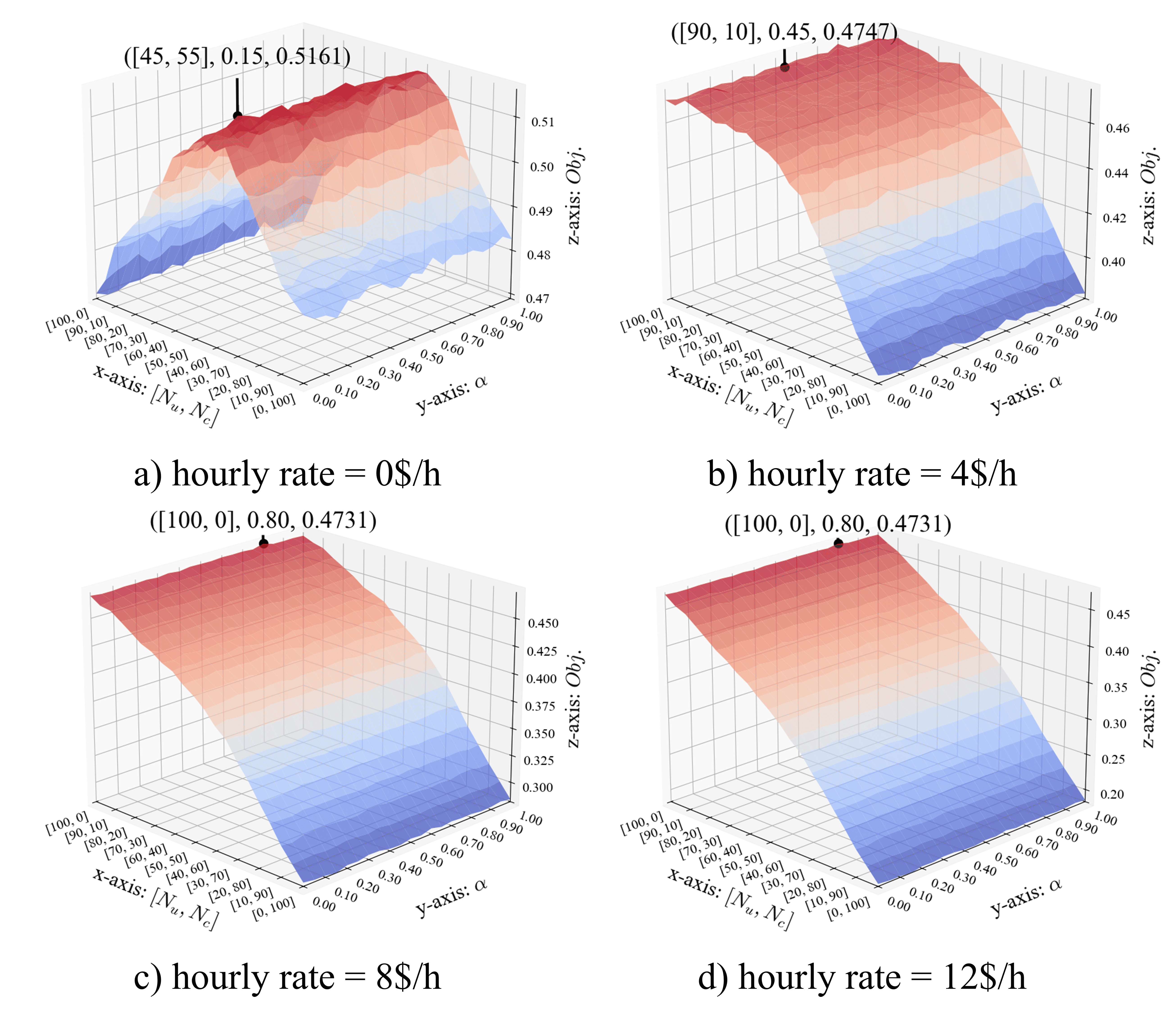}
    \caption{Objective value improvements for varying hourly rates with $k=0.2$.}
    \label{fig:result_optimization_k=0.2}
\end{figure}

\begin{figure}[t]
    \centering
    \includegraphics[width=0.50\linewidth]{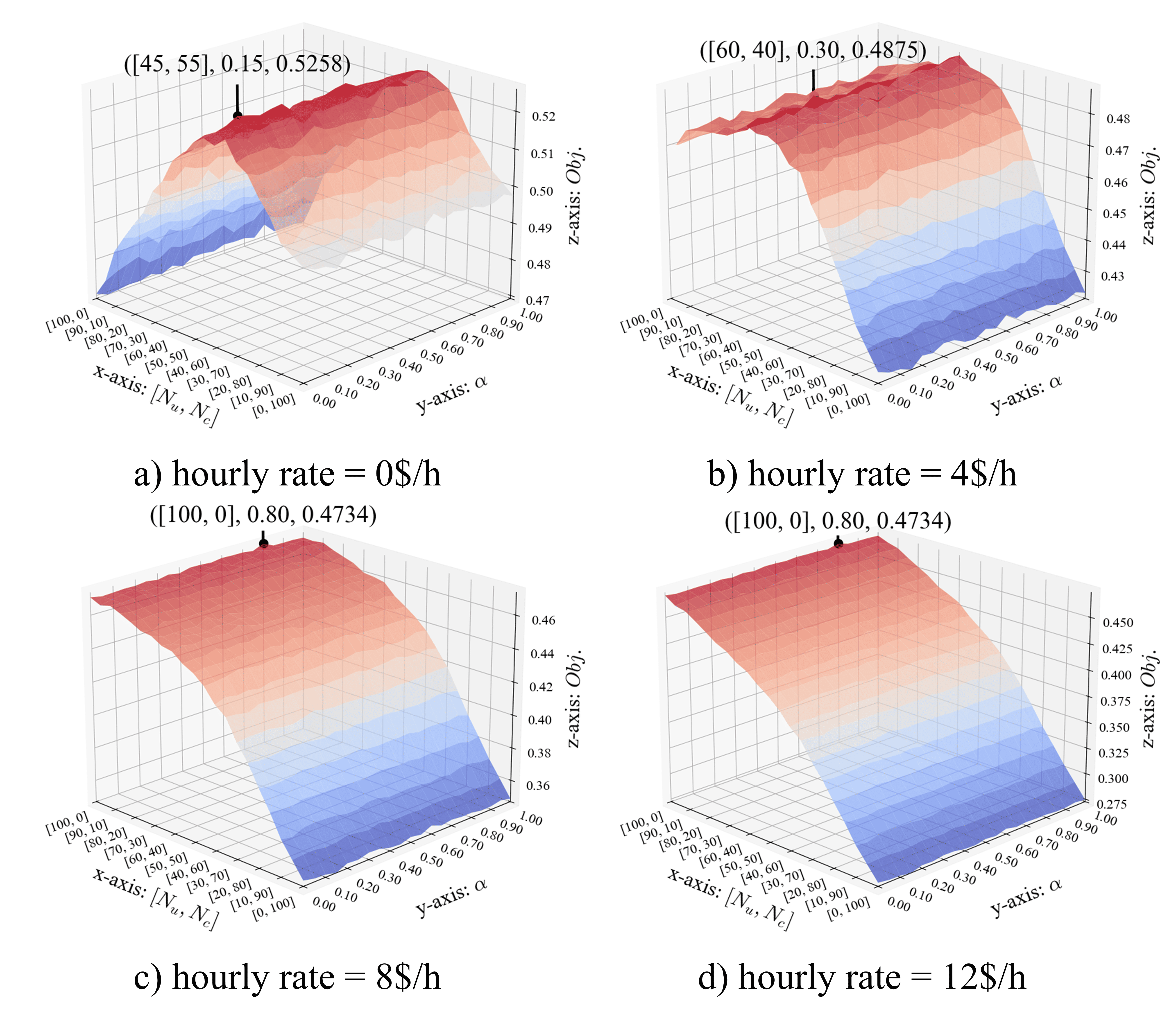}
    \caption{Objective value improvements for varying hourly rates with $k=0.4$.}
    \label{fig:result_optimization_k=0.4}
\end{figure}

\begin{figure}[t]
    \centering
    \includegraphics[width=0.50\linewidth]{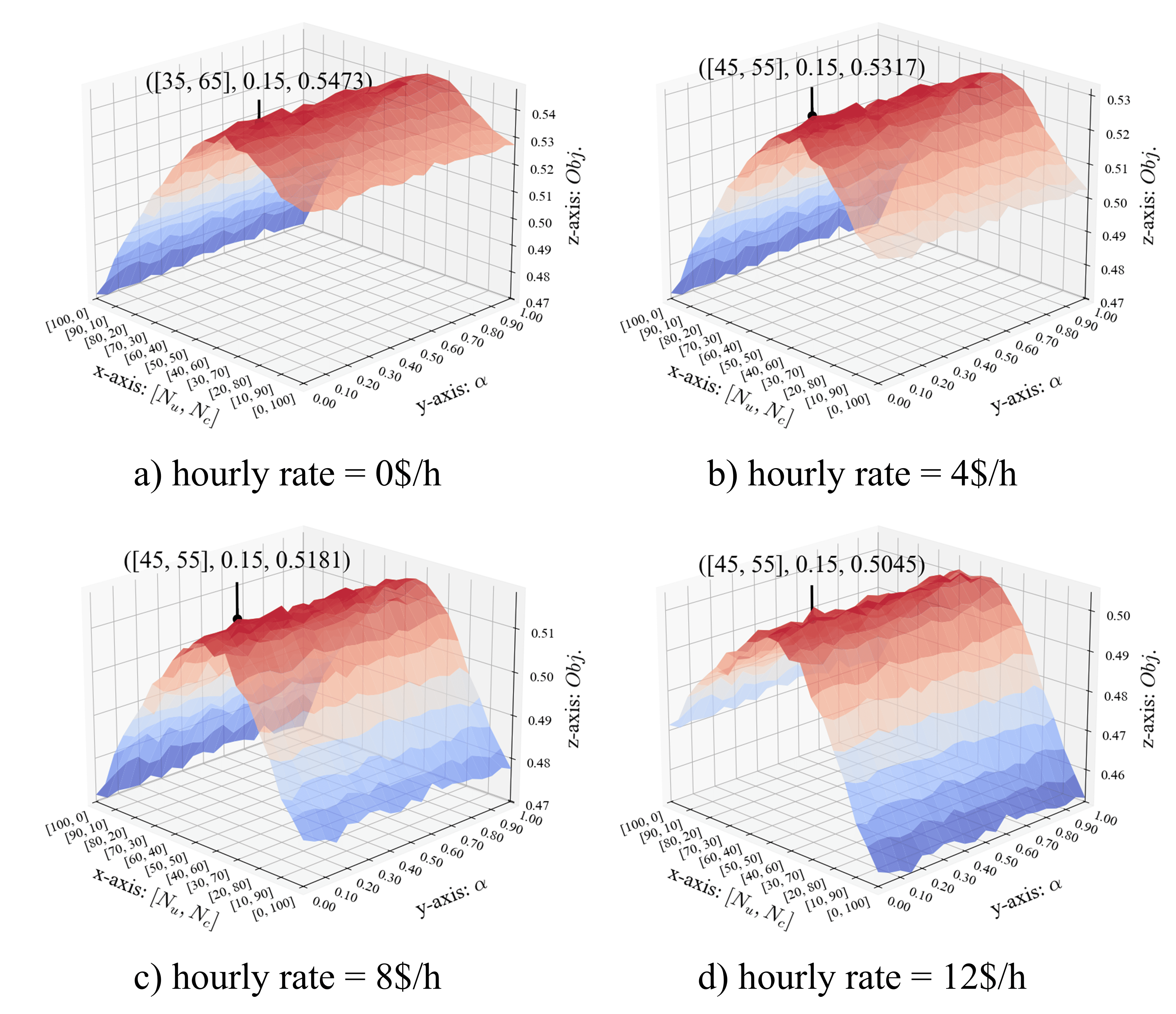}
    \caption{Objective value improvements for varying hourly rates with $k=0.8$.}
    \label{fig:result_optimization_k=0.8}
\end{figure}

The results for different hourly rates when $k$=0.2, 0.4, and 0.8 are presented in Figure~\ref{fig:result_optimization_k=0.2}, \ref{fig:result_optimization_k=0.4}, and \ref{fig:result_optimization_k=0.8}.

\subsection{Training hypernetworks: Entire vs. Segments}\label{app_subsec:exp_segment}

The experiment aimed to investigate whether training hypernetworks specialized for specific segments of the design space would yield better performance compared to training a single hypernetwork over the entire design space. The objective function followed Equation~(\ref{eq:objective_function}), with $k=0.6$ and an hourly rate set at 8\$/h. Two training approaches were compared: one involving the entire design space $(N_c \in [0,100], \alpha \in [0,1])$ and another involving four distinct segments of the design space: $([0,50], [0,0.5])$, $([0,50], [0.5,1.0])$, $([50,100], [0,0.5])$, and $([50,100], [0,0.5])$. The former approach trained a single hypernetwork over 60,000 episodes, while the latter trained a specialized hypernetwork for each segment over 15,000 episodes per segment.

\begin{figure}[t]
    \centering
    \includegraphics[width=0.50\linewidth]{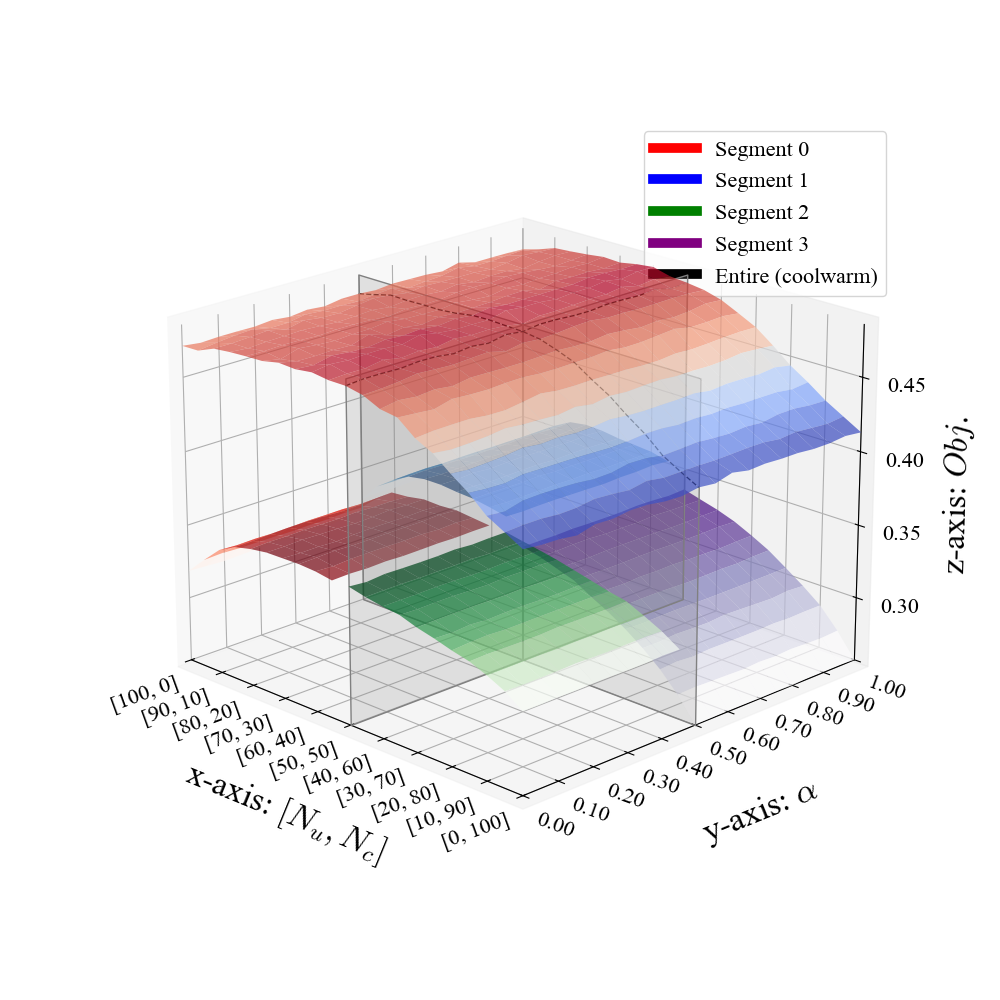}
    \caption{Comparison between a single hypernetwork over the entire design space and hypernetworks for each segment.}
    \label{fig:result_segments}
\end{figure}

Figure~\ref{fig:result_segments} demonstrated that training a single hypernetwork over the entire design space yielded better performance compared to training specialized hypernetworks for individual segments when using the same total training budget. This outcome suggests that a hypernetwork trained on the entire design space benefits from information sharing across similar configurations, enhancing its generalization and efficiency. 

%%%%%%%%%%%%%%%%%%%%%%%%%%%%%%%%%%%%%%%%%%%%%%%%%%%%%%%%%%%%

\end{document}